\documentclass[]{aa}
\usepackage{natbib}
\usepackage{graphicx}
\usepackage{wasysym}
\usepackage{txfonts}
\usepackage{color}
\def\HH{\mbox{H$_2$}}

\def\fH2{\mbox{f$_\HH$}}

\def\EBV{\mbox{E$_{\rm B-V}$}}

\def\nH2{\mbox{${\rm n}_\HH}$}

\def\pccc{~{\rm cm}^{-3}} 
 
\def\pcc{~{\rm cm}^{-2}}

\def\Tsub#1 {\mbox{${\rm T}_{\rm #1}$}}
\def\TK  {\Tsub K }

\def\arcsec{\mbox{$^{\prime\prime}$}} \def\arcmin{\mbox{$^{\prime}$}}

\def\degr{$^{\rm o}$}
\def\p{\mbox{$^+$}}

\def\XCOa{\mbox{X$^0_{\rm CO}$}}
\def\WCOa{\mbox{W$^0_{\rm CO}$}}

\def\cch{\mbox{C$_2$H}}

\def\h13cop{\mbox{{H$^{13}$CO\p}}}

\def\C3H{\mbox{C$_3$H}}
\def\c3h2{\mbox{C$_3$H$_2$}}
\def\cc3h2{\mbox{{\it c}-C$_3$H$_2$}}
 
 \def\R0{R$_0$}
\def\G0{\mbox{G$_0$}}

\def\ddeg{{}^\circ\kern-.1em}

\def \kms{\mbox{km\,s$^{-1}$}}

\def\E#1 {$10^{#1}$}
\def\E#1 {E{#1}}
\def\P#1,{$\nH2\TK~=~#1\times~10^4\pccc$~K}
\def\ec#1,#2,#3,{#1\,(#2)\E{#3}}

\def\H3{\mbox{H$_3$}}

\def\RH2{\mbox{R$_{\rm G}$}}
\def\g13{\mbox{g$_{13}$}} 

\def\cc3h{\mbox{{\it c}-\C3H}}
\def\lc3h{\mbox{{\it l}-\C3H}}

\newcommand{\emm}[1]{\ensuremath{#1}}   
\newcommand{\emr}[1]{\emm{\mathrm{#1}}} 

\newcommand{\hcop}{\emr{HCO^+}}

\renewcommand{\coth}{\emr{^{13}CO}}
\newcommand{\coei}{\emr{C^{18}O}}

\newcommand{\X}[1]{\emm{X_\emr{#1}}}

\newcommand{\XCO}{\X{CO}}

\newcommand{\W}[1]{\emm{{\rm W}_\emr{#1}}}
\newcommand{\WCO}{\W{CO}}

\title{Molecular Gas and Dark Neutral Medium in the Outskirts of Chamaeleon}

\author{ H. Liszt\inst{1} and M. Gerin\inst{2} and I. Grenier\inst{3}}
\institute{
     National Radio Astronomy Observatory,
           520 Edgemont Road,
           Charlottesville, VA,
           USA 22903 
      \email{hliszt@nrao.edu}
\and
LERMA, Observatoire de Paris,  PSL Research University, CNRS
Sorbonne Universit\'es, UPMC Univ. Paris 06, Ecole Normale
  Sup\'erieure, F-75005 Paris, France
\email{maryvonne.gerin@ens.fr}
\and
AIM, 
CEA-IRFU/CNRS/Universit\'e Paris Diderot,
D\'epartement d'Astrophysique, CEA,
Saclay, 91191, Gif-sur-Yvette, France
\email{isabelle.grenier@cea.fr}
}
\begin{document}
\date{received \today}%
\offprints{H. S. Liszt}%
\mail{hliszt@nrao.edu}%
%
\abstract
{More gas is  inferred to be present in molecular cloud complexes 
 than can be accounted for by HI  and CO emission, a phenomenon known as 
 dark neutral medium (DNM) or CO-dark gas for the molecular part.}
{To see if molecular gas can be detected in Chamaeleon
where gas column densities in the DNM were inferred and CO emission was 
 not detected.}
{We took $\lambda$3mm absorption profiles of \hcop\ and other molecules 
 toward thirteen background quasars across the Chamaeleon complex, only one of 
 which had detectable CO emission.  We derived 
 the \HH\ column density assuming N(\hcop)/N(\HH)$ = 3\times 10^{-9}$ as before.}
 {With the possible exception of one weak continuum target, \hcop\ absorption 
 was detected in all directions, \cch\ in 8 directions and HCN in 4 
 directions.  The sightlines divide in two groups according to 
 their DNM content with one group
 of eight directions having N(DNM) $ \ga 2\times10^{20} \pcc$ and  
 another group of five directions having N(DNM) $< 0.5\times10^{20}\pcc$.
 The groups have comparable mean N(HI) associated with Chamaeleon
  $6-7 \times 10^{20}\pcc$ and total hydrogen column density per unit 
 reddening $6-7 \times 10^{21}\pcc$ mag$^{-1}$.  They differ, however,
 in having  quite different  mean reddening 0.33 vs 0.18 mag, mean N(DNM) 
 3.3 vs 0.14 $\times 10^{20}\pcc$ and mean molecular column density
 2N(\HH) = 5.6 vs 0.8 $\times 10^{20} \pcc$.  The gas at
 more positive velocities is enriched in molecules and DNM.}
 {Overall the quantity of \HH\ inferred from \hcop\ can fully account
 for the previously-inferred DNM along the sightlines studied here.  
 \HH\ is concentrated 
 in the high-DNM group, where the molecular fraction is 46\% vs. 13\%
 otherwise and 38\% overall. Thus, neutral gas in the outskirts of 
 the complex is mostly atomic but the DNM is mostly molecular. 
 Saturation of the HI emission line profile may occur along 3 of
 the 4 sightlines having the largest DNM column densities
 but there is no substantial reservoir of ``dark'' 
 atomic or molecular gas that remains undetected as part of the 
 inventory of dark neutral medium.}


\authorrunning{G$^2$L} \titlerunning{Chamaeleon's dark and molecular gas}

\maketitle{}

%

\section{Introduction}

The use of  $\lambda 21$cm H I and $\lambda 2.6$mm CO emission to trace 
the column densities of atomic and molecular hydrogen is fundamental to 
study of the interstellar medium (ISM).  Yet, it is commonly the case that
a combination of these two tracers fails to account for all of the gas
that is inferred to be present from other tracers of the total column density,
for instance the Fermi $\gamma$-ray flux \citep{GreCas+05} and the Planck 
sub-mm dust optical depth  \citep{Pla15Cham}.  This has led to the
concept of ``dark'' gas \citep{GreCas+05} which could, in principle and 
depending
on particular circumstances, be attributed either to saturation of the
H I emission profile \citep{FukTor+15,OkaYam+17} or to the inability of
CO emission to track the \HH\ column density at intermediate reddening
 \citep{wolfire10,Pla15Cham,RemGre+18} where the column density of \HH\
is appreciable but N(CO) is less than $ 10^{15}\pcc$ and the integrated CO 
brightness \WCO\ is below 1 K-\kms\ \citep{Lis17CO}.  \cite{DonMag17} argue 
that the CO-dark gas 
inventory in Pegasus-Pisces is halved by their extensive new detections 
of CO emission at levels \WCO\ $<$ 1 K-\kms.

A flexible approach to determining the origin of the missing gas is to 
express the total gas column density N(H) as a linear 
combination of N(H I) $\propto$ W$_{\rm H I}$ and N(\HH) $\propto$ \WCO\
as represented by the H I and CO emission profile integrals 
W$_{\rm H I}$ and \WCO, but with an added term N(DNM) for the so-called 
dark neutral medium that is absent in emission and could be in either or
both atomic and  molecular form \citep{Pla15Cham,RemGre+18}. 
By comparing the linear combination 
N(H)$_{\rm emp}$ = N(H I) + N(DNM)+ 2N(\HH) with the map of a (total) 
column density tracer such as the 
gamma-ray flux or dust opacity, the spatial distribution of the DNM  
 can be derived over an entire cloud complex, thereby 
associating the DNM  with the distribution of the 
H I and CO tracers and with the structure of the cloud complex itself. 

In this work we take advantage of an existing analysis of the DNM in 
the Chamaeleon cloud complex \citep{Pla15Cham} to compare DNM and 
molecular gas, but with an \HH\ tracer that is sensitive to very weakly-excited
molecular gas and to molecular gas having smaller column density than
is typically sampled in CO emission:  we observed \hcop\ and other 
strongly-polar species in absorption at $\lambda$3mm toward 13 relatively
strong background continuum sources seen by chance against the outskirts 
of the Chamaeleon complex where CO had been detected in only one direction 
with \WCO\ = 2.4 K-\kms, only slightly above the detection limit.

In this work  we discuss the implications of our detection of molecular 
absorption in all of the directions lacking observable CO emission over 
wide areas of the outskirts of the Chamaeleon cloud complex.  Sect. 2 presents
the new observations, summarizes existing observational material that 
is cited and the underlying assumptions that undergird our work.  
In Sect. 3 we discuss the new results and Sect. 4 is a summary.

\section{Observations and data reduction}

\begin{figure*}
\includegraphics[height=19cm,angle=90]{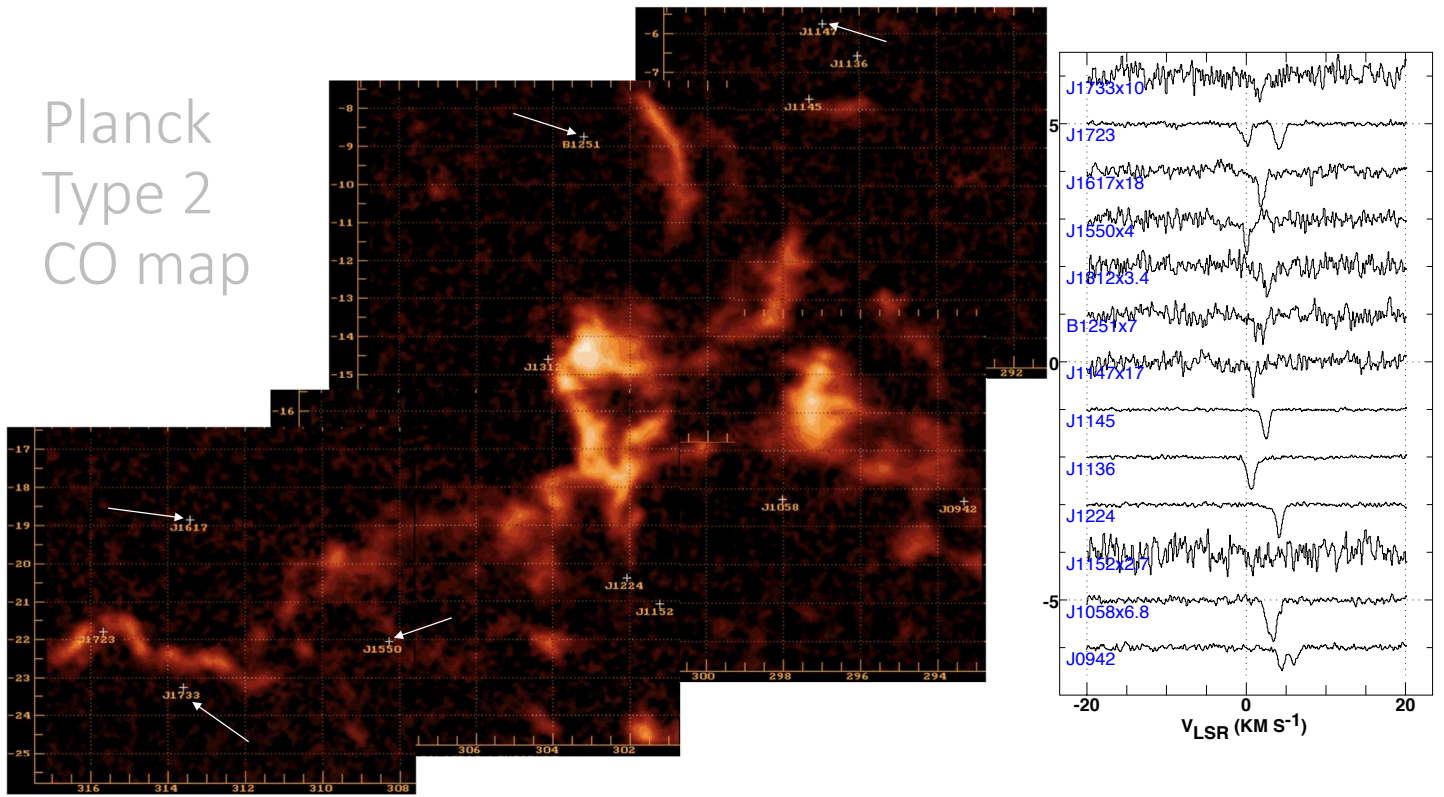}
  \caption[]{Left: The positions of the background sources observed here
are projected against a mosaic of cutouts of Planck Type 2 integrated
 CO emission: the color scale runs from 0 to 23 K-\kms.  The five sources
comprising the group with small N(DNM) are noted with arrows. Right: the \hcop\
absorption profiles in all directions, shifted vertically and scaled up 
in some cases.}
\end{figure*}

\begin{figure*}
\includegraphics[height=11.7cm]{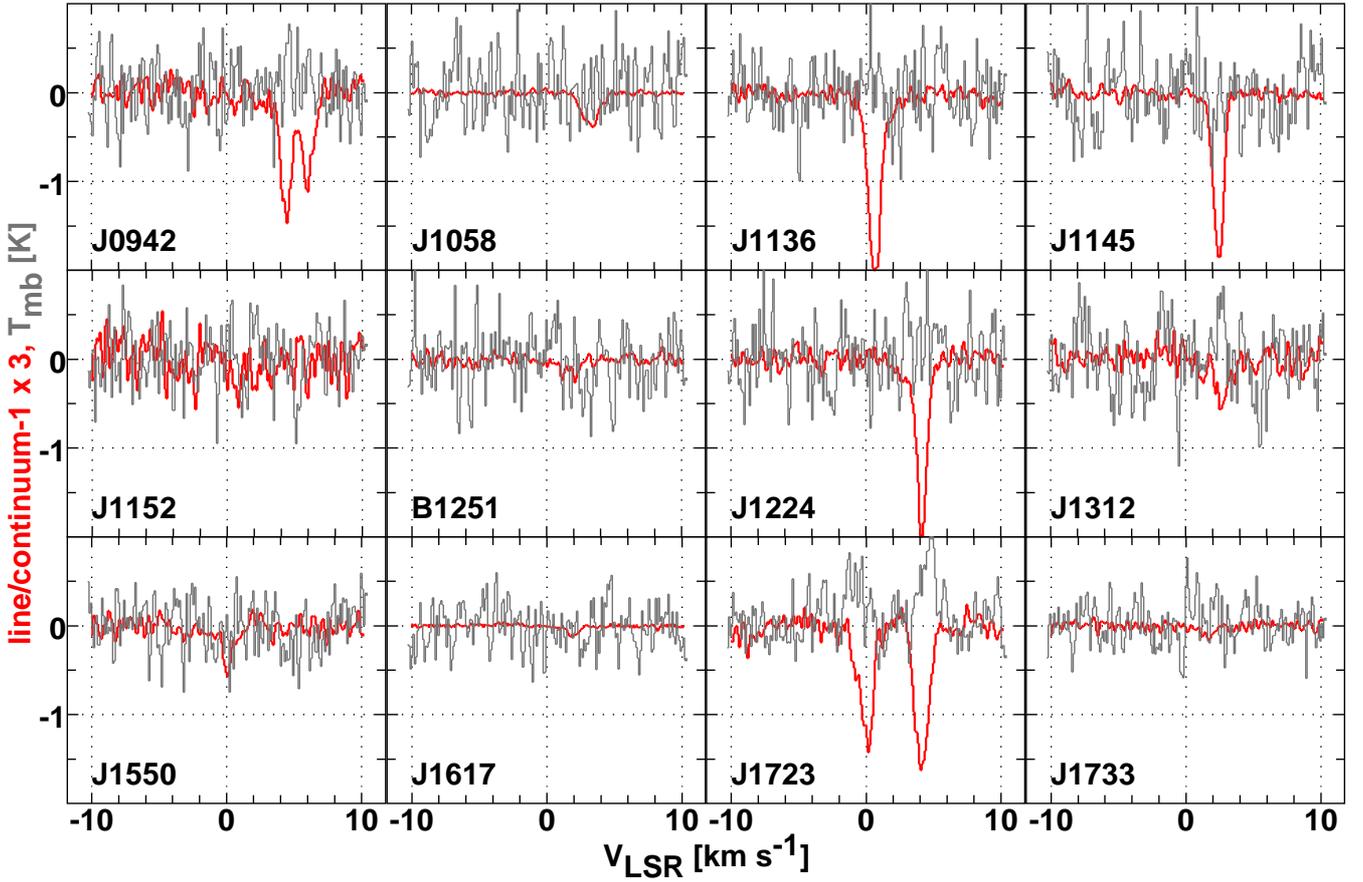}
  \caption[]{Comparison of NANTEN CO emission (black) and ALMA \hcop\ absorption 
(red) profiles in the directions observed. The CO emission is shown on the 
main-beam scale of the NANTEN telescope in Kelvins and the \hcop\ absorption 
profiles have been scaled up by a factor three. CO was reliably detected 
only toward J1723 and not observed toward J1147 (not shown here).}
\end{figure*}

\begin{figure*}
\includegraphics[height=9cm]{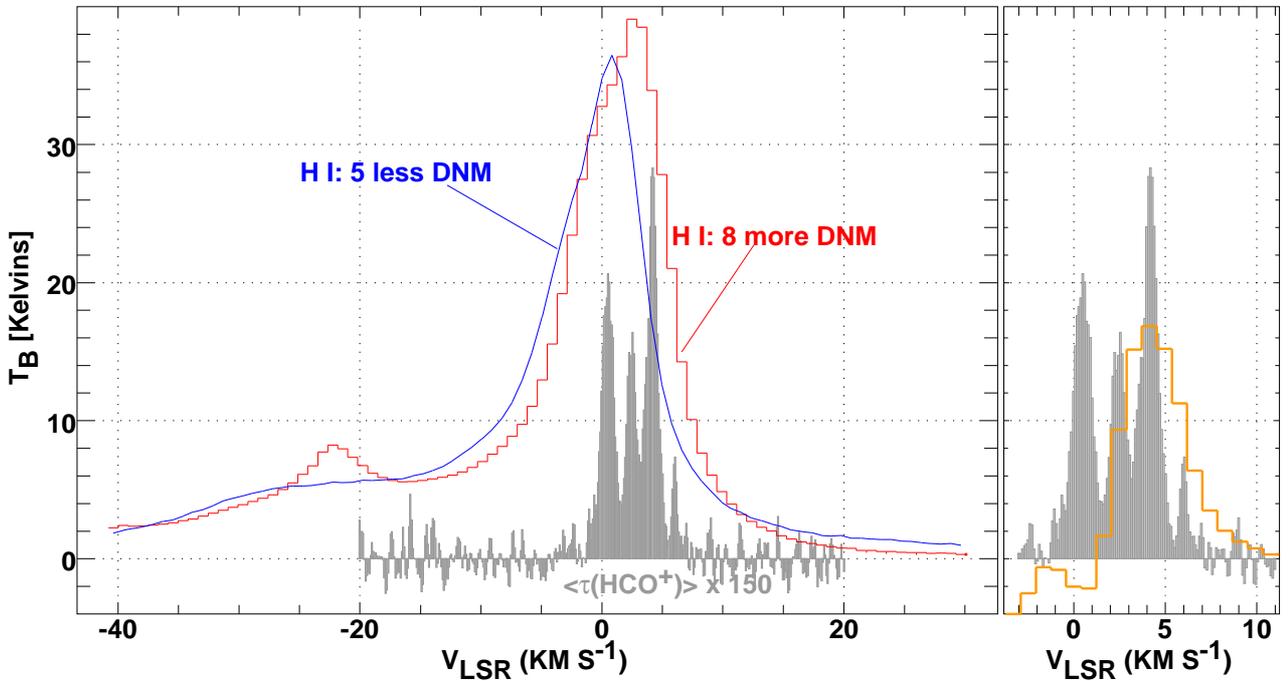}
\caption[]{Kinematics of the H I, separated by N(DNM). Left: 
Shown separately are the mean H I profiles for the two groups of sources 
according to their N(DNM): H I emission in the group having more DNM is 
plotted as a red histogram and emission in the other group is shown as a blue line.
Also shown is the mean \hcop\ optical depth profile for all sources 
scaled upward by a factor 150. Right: as at left but the H I profile 
plotted in orange is the 
difference of the H I profiles shown at left.  The greatest difference between 
the H I profiles coincides with the peak of the \hcop\ optical depth.}
\end{figure*}
 
\begin{figure}
\includegraphics[height=7.3cm]{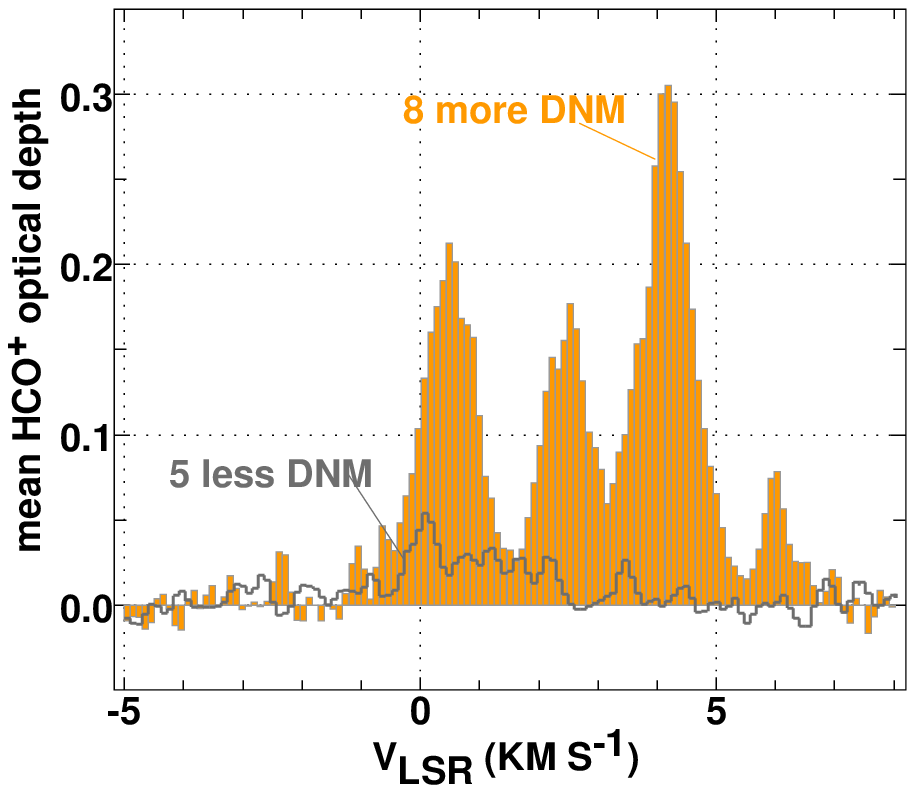}
 \caption[]{Kinematics of the molecular gas traced by \hcop, separated by 
N(DNM).  Shown separately are the average \hcop\ optical depths of the five 
 sources with weak NDNM (dark gray histogram) and the eight sources with 
stronger N(DNM)(orange bars).}
\end{figure}

\subsection{New ALMA absorption measurements and \HH\ column densities}

We observed the J=1-0 lines of \hcop, HCN and \cch\ in absorption toward the 
13 continuum sources listed in Table 1, whose locations on the sky are
shown in Fig. 1. CS J=2-1 and H$^{13}$CO\p\ were observed
but not detected and are not discussed in the text below. 
The
work was conducted  under ALMA Cycle 4 project 2016.1.00714.S whose pipeline data 
products were delivered in 2017 February and March. Spectra were extracted from the 
continuum-subtracted pipeline-processed data cubes at the 
pixel of peak continuum flux in the continuum map
made from the spectral window containing each line, and divided by the 
continuum flux in the continuum map for that spectral window at that pixel.  
Fluxes at 89.2 GHz ranged from 0.11 Jy for J1152 to 2.5 Jy for J1617 as noted 
in Table 1.  Each spectrum consisted of 1919 semi-independent channels spaced 
61 kHz corresponding to 0.205 \kms\ at the 89.189 GHz rest frequency of 
\hcop.  Absorption from  CS and 
H$^{13}$CO\p\ was not detected and will not be further discussed. 

\HH\ column densities N(\HH) are derived from the absorption line profile
of \hcop\ using N(\hcop) $ = 1.10 \times 10^{12} \pcc \int \tau~{\rm dv}$ 
with the profile integral in units of \kms\ \citep{AndKoh+16,GerLis17}.
N(\hcop) is derived under the assumption that the rotational excitation 
temperature is that of the cosmic microwave background, which is justified 
by the weakness of \hcop\
emission in diffuse molecular gas, even when the absorption is quite optically
thick \citep{Lis12,LisPet16}.  We also assume N(\hcop)/N(\HH) $= 3\times 10^{-9}$.
The relative abundance N(\hcop)/N(\HH) can be derived in various ways; 
most direct is to refer it to the relative abundances of OH and CH 
\citep{LisLuc96,LisGer16}, both of which are directly determined optically, 
eg \citep{WesGal+10}, and give the same value for N(\hcop)/N(\HH).   

\subsection{H I column densities}

H I column densities were derived for all directions from Galactic All-Sky Survey 
GASS III $\lambda$21 cm H I emission spectra \citep{KalHau15}.  Table 1 shows total
column densities N(H I) derived from the line brightness profile integral 
over all velocities in the optically thin limit, 
$$ N(H I)|_{tot} = 1.823 \times 10^{18} \pcc \int T_B~{\rm dv}. $$
Column densities of H I associated with the Chamaeleon complex, 
N(H I)$|_{\rm cham}$, are given in Table 2.  These were derived following 
the methodology described by  \cite{Pla15Cham} and \cite{RemGre+18}.  
GASS III H I emission profiles were decomposed into pseudo-Voigt profiles 
(gaussian cores with the possibility of co-centered Lorentzian wings) and the 
profile integrals of all components centered within the velocity range of the 
Chamaeleon complex were summed, again in the optically thin limit that 
provided the best fit in the analysis of \cite{Pla15Cham}.  


\subsection{DNM}

Maps of the column density of dark neutral medium - the component that is not 
fully visible in H I or CO emission - were published for most of the sightlines 
by \cite{Pla15Cham}.  Specific values of N(DNM) along the present sightlines 
were extracted from that work for the present use, along with values toward 
J1136 and J1147 that were just outside the region considered by \cite{Pla15Cham}.  
for their interstellar fit but still in the analysis region.  All of these values 
are given in Table 2.   N(DNM) is comparatively small in 5 directions and no DNM 
was needed along three of these: a determination of N(DMN) was made along all 
sightlines including those for which 0-values are given in Table 2.

\subsection{CO emission}

The derivation of N(DNM) by \cite{Pla15Cham} used NANTEN maps of J=1-0 CO 
emission from \cite{MizYam+01}.  The CO emission map in Fig. 1 uses the 
Planck Type 2 CO product which is virtually indistinguishable from the 
NANTEN map in a blink comparison. The rms of the Planck Type 2 CO product 
over the fields comprising Fig. 1 is 
$\Delta$\WCO\ $= 0.45\pm0.05$ K-\kms, leading to a detection limit comparable 
to that of the NANTEN maps, about 2 K-\kms\ \citep{MizYam+01}. It should be noted 
that the Planck CO maps have been hashed over small regions in the immediate 
vicinity of several of our sightlines, presumably as the result of
point-source removal, something of a Catch-22 situation for the present
study.

\subsection{Reddening and dust optical depth}

The 6\arcmin\ resolution dust-emission maps scaled to optical
reddening \EBV\ by \cite{SchFin+98} are cited in Tables 1 and 2. Those 
reddening values can be converted to Planck 353 GHz dust optical 
depth using the relationship established by \cite{PlaXI} between
the 353 GHz dust optical depth and the \EBV\ values of \cite{SchFin+98}, 
\EBV/$\tau_{353} = (1.49\pm0.03) \times 10^4$ mag.

\subsection{Conventions}

Velocities presented with the spectra are taken with respect to the kinematic 
definition of the Local Standard of Rest.  N(H) is the column 
density of H-nuclei in neutral atomic and molecular form, N(H) = N(H I)+2N(\HH).

\section{Molecules and Dark Neutral Medium in the Chamaeleon Complex}

The use of H I, CO, Planck sub-mm dust and Fermi $\gamma$-ray emission
observed on 4\arcmin - 16\arcmin\ and larger angular scales to derive  
gas properties across the Chamaeleon complex was discussed by \cite{Pla15Cham} 
and further refined by \cite{RemGre+18}.  This work adds the information 
gained from molecular absorption profiles observed along thirteen 
sightlines toward point-like  extragalactic mm-wave continuum sources that 
otherwise serve as phase calibrators for the ALMA telescope. 

The angular resolution of interferometric absorption line observations is the 
apparent angular size of the background source in the presence of interstellar 
scattering, which is milli-arcseconds at radio wavelengths: the instrumental 
resolution, approximately 1\arcsec\ during the observing, does not determine this.  
Broader-beam emission observations do more spatial averaging, but both types 
of observations do comparable sampling of the gas along the line of sight which 
is the more important aspect.  Differences between emission and absorption 
line profiles generally reflect the character of the gas distribution, 
not the differing spatial resolution that was employed to obtain them.  
\citep{Cla65}.

This work takes advantage of the fact that \hcop\ absorption is more 
widespread than CO emission 
\citep{LucLis96,LisLuc98,LisPet12}, presumably because N(CO) varies 
so sharply with N(\HH) and \EBV\ \citep{BurFra+07,SonWel+07,SheRog+08} 
while N(\hcop)/N(\HH) can be taken as constant.  The net result is that 
\hcop\ absorption is the more reliable tracer of \HH\ along the lines of sight 
observed here and even very extreme differences in angular resolution may 
introduce scatter, but  do not introduce bias.  To accommodate this, 
we derive and tabulate the 
properties of the gas along the individual sightlines (Tables 1 and 2), as we must, 
but the discussion is cast in terms of the average properties of two groups and 
the differences between them.  Discussion of individual sightlines is largely left 
for the concluding parts of the discussion and the final figure.

\subsection{General properties}

General properties of the sightlines studied here are given in Table 1, with
entries as explained in Sect. 2. A finding chart is presented in Fig. 1, where
the locations of the continuum targets are shown against a background image 
of the Planck Type 2 CO map and the 5 sightlines having small column 
densities N(DNM) $ < 0.5 \times 10^{20} \pcc$ are noted.  All of the \hcop\ 
absorption profiles are shown at right in Fig. 1, in some cases scaled up 
for easier viewing: they are shown  again in Fig. 2 with the NANTEN 
CO profiles (see Sect. 3.2).  

The continuum sources sample the borders of the CO emission distribution
and avoid the regions of detected CO emission, except for J1723.
A recent search of the ALMA calibrator database for newly-discovered strong 
continuum background targets in Chamaeleon was not fruitful.  Unless weaker 
continuum sources are used, the current survey represents the state of the 
art in this part of the sky.

In Table 2 the sources are reordered by descending N(DNM), and grouped 
according to whether N(DNM) $\ga 2\times 10^{20}~{\rm H}~\pcc$ (8 sources) 
or N(DNM) $< 0.5\times 10^{20}~{\rm H}~\pcc$ (5 sources). The sightlines
with smaller N(DNM) are called out in Fig. 1 where it is seen that they
occur across the entire length of the Chamaeleon cloud complex.  This is
noteworthy, given the limited kinematic range of the molecular absorption
in these directions, see Sect. 3.2.

The two groups have similar mean total H I column density 
$<$N(H I)$|_{\rm tot}> = 13\times 10^{20}\pcc$ averaged over all
sightlines in each group, similar mean N(H I) associated 
with Chamaeleon $<$N(H I)$|_{\rm cham}> = 5.6-6.6\times 10^{20}\pcc$ and 
similar mean total column density per unit reddening
$<$N(H)/\EBV $>$ $\approx 5.4-6.7 \times 10^{21} \pcc$ (mag)$^{-1}$, but the 
group of eight DNM-rich sightlines has much larger mean reddening 
$<$\EBV$>$ (0.33 vs. 0.18 mag), mean DNM column density
$<$N(DNM) $>$ (3.3 vs. 0.14 $\times 10^{20} \pcc$) and column density of 
H-nuclei in molecular form $<$2N(\HH)$>$ (5.6 vs 0.83 $\times 10^{20} \pcc$). 
Their similarity in terms of the total column density per unit reddening 
is only possible because of the larger contribution of \HH\ in the group 
with higher mean reddening.

\subsection{Kinematics and abundance}

The kinematic signature of the difference between the two groups is illustrated 
in Fig. 3 where the mean H I profiles for each group (at left) and their 
difference (at right) are plotted along with the mean \hcop\ optical depth
profile, suitably scaled.  The figure makes the point that the difference
in H I, an increase, occurs at 2 - 8 \kms\ where most of the molecular gas is 
found.  The low-velocity component of \hcop\ at v $\approx$ 0.5 \kms\ is quite 
widespread, being present toward J1136, J1147, J1550 and J1723 and is
present in both DNM-groups.

This point is further elaborated in Fig. 4, comparing the mean \hcop\ optical
depth profiles in the two groups.  \hcop\ absorption in the group with 
lower N(DNM) is of course weaker, but it also occurs only at v $\la 2$ \kms.  
As shown in Fig. 1,  sightlines with small N(DNM) span the breadth
of the Chamaeleon complex and in principle could sample the full range of 
velocities, but they do not.  This  suggests that the higher molecular abundance
and increased kinematic complexity in the higher DNM group are associated,
as for instance when \HH\ production is enhanced in the presence of slow
shocks associated with turbulent flows \citep{MicGlo+12,ValHen+18} 

Figure 5 compares the mean optical depth profiles of \hcop, HCN and \cch.
The HCN-\hcop\ relationship is monotonic in the sense that the weakest and 
strongest components are the same in both species but the HCN/\hcop\ ratio 
 varies and is noticeably larger in the strongest component at 4 \kms. 
This mimics the N(HCN)-N(\hcop) relationship that is seen along individual
lines of sight where there is  a rapid increase of N(HCN) at
N(\hcop) $\ga 10^{12}\pcc$ \citep{LisLuc01,AndKoh+16,RiqBro+18,LisGer18}
or N(\HH) $\ga 3\times 10^{20} \pcc$.  N(HCN) and N(CN) are seen to vary in 
fixed proportion and CN is well-known 
to increase in abundance for N(\HH) $\ga 3\times 10^{20} \pcc$ in optical 
spectroscopy of the diffuse molecular gas \citep{SheRog+08}.  
Although \cch\ absorption is faint, the \cch\ and \hcop\ line profiles 
look fairly similar.  \cch\ is known to be present in the diffuse molecular 
gas in quasi-constant abundance ratio relative to CH and to \hcop. The 
present data are consistent with the established relation between 
\hcop\ and \cch\ \citep{LucLis00C2H,AndKoh+16,RiqBro+18,LisGer18}.

\begin{figure}
\includegraphics[height=7.3cm]{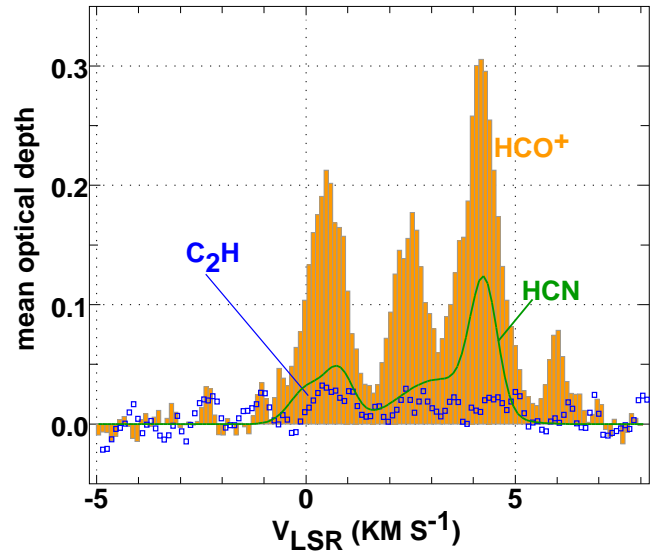}
  \caption[]{Mean optical depth profiles for strongly-polar molecules
for the eight sources with N(DNM) $ \ga 2 \times 10^{20}\pcc$. The HCN
profile in green shows the kinematic structure that would be convolved 
with the HCN hyperfine splitting to reproduce the observations assuming
that the hyperfine components appear in the LTE ratio 5:3:1. 
\cch\ is shown in blue.}
\end{figure}


\subsection{Balance among N(\HH), N(DNM) and N(H I)}

Overall, the hydrogen in the outskirts of the Chamaeleon complex
is mainly in atomic form.  In ascending order, averaged over all sightlines, 
$<$N(DNM)$> = 2.1\times10^{20}\pcc$, 2N(\HH) $= 3.8\times10^{20}\pcc $
$<$N(H I)$|_{\rm cham}> = 6.2\times10^{20}\pcc $
and this ordering applies to most of the sightlines individually as well.
There is only one direction (J1152) where N(DNM) even slightly exceeds 
N(H I)$|_{\rm cham}$ and only three where 2N(\HH) $>$ N(H I)$|_{\rm cham}$.
The point is that the molecular component must represent a minority fraction 
of the total hydrogen even if all the DNM turns out to be \HH.  As summarized 
in Table 3, \HH\ comprises 46\% of the 
H-nuclei in the group having more DNM, but only 13\% otherwise and 36\%,
about one-third, in total.  Overall, the molecular/atomic balance of the 
outlying Chamaeleon gas is very similar to that of the diffuse ISM as a 
whole, with estimates of the hydrogen fraction in \HH\ ranging from 25 to 
40 percent \citep{SavDra+77,LisLuc02}.

\subsection{CO emission, observed and expected}

\begin{figure}
\includegraphics[height=13.5cm]{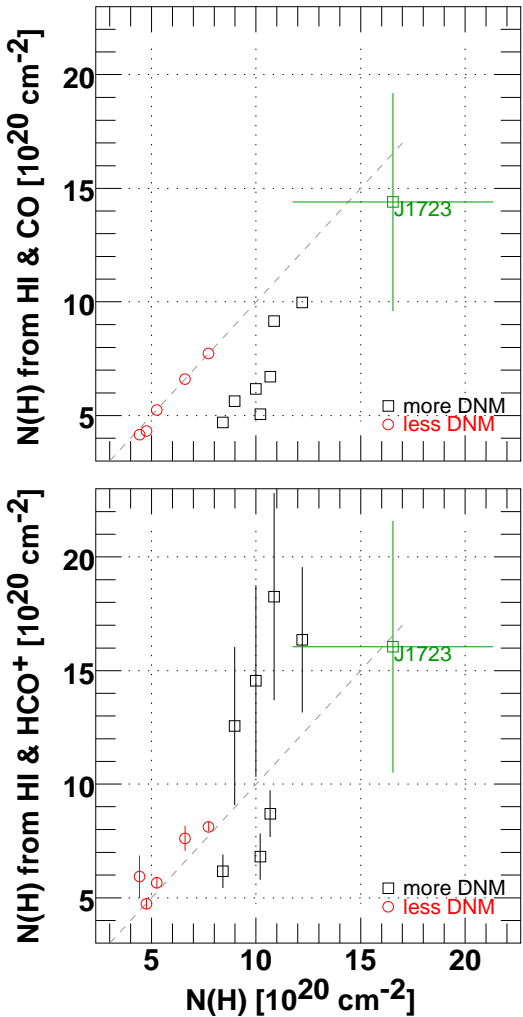}
  \caption[]{Comparison of total column density of H-nuclei N(H) derived from 
the DNM analysis and from observable column densities.  The quantity plotted 
along the horizontal axis, N(H) = N(H I)$|_{\rm cham}$+N(DNM)+2\WCO \XCOa, is 
the quantity fitted in the global DNM analysis and is the same in both panels.
The vertical axis in each panel is the inferred total hydrogen column density 
N(H) = N(H I)$|_{\rm cham}$+2N(\HH) but calculated differently: at top, with 
N(\HH) = \WCO \XCOa, and at bottom with N(\HH) = N(\hcop)$/3\times10^{-9}$.
The error bars assume $\pm50$\% errors in \WCO\ and N(\hcop)/N(\HH).
}
\end{figure}

The NANTEN CO profiles toward our targets are shown and compared with the 
ALMA \hcop\ absorption profiles in Fig. 2, except toward J1147 that was
not covered in the CO mapping.  CO emission was reliably detected 
only toward J1723 where the CO and \hcop\ lines coincide in velocity 
and \WCO\ = 2.4 K-\kms. The CO profile toward J0942 is suggestive, 
given the overlap with \hcop\ absorption, but other equally strong and 
apparently spurious features are present in the CO spectrum. 
Weak spurious signals in consecutive channels are
seen in many CO spectra. For example, toward J1312 there are three 
equally strong positive-going features, only one of which coincides with the \hcop\ 
absorption, and three comparably strong negative-going features.  
Moreover the \hcop\ absorption toward J1312 is too weak to support a 
credible CO detection, as is also the case for J1152. For J1224 a weak 
positive-going CO feature near 0-velocity is too far removed from the 
\hcop\ absorption to be reliable.

A null CO signal, \WCO\ = 0, was therefore taken in all directions except 
J1723 in the succeeding discussion.  In this regard, note that we hedged 
this assumption with an accepted ALMA Cycle 5 proposal to search for CO absorption
so that our assumption of null detections will be tested.  Perhaps more
importantly, the main conclusions of this work do not stand to be affected
by overlooking a few weak detections of CO emission.  The corresponding 
amounts of molecular hydrogen are small compared to the quantity of DNM,
and small compared to the amounts of \HH\ that was inferred from the 
\hcop\ absorption.

The sightlines toward J1723 and J0942 have the highest molecular column 
density and would be expected to have the strongest CO emission.
A typical global average for the conversion from N(\HH) to \WCO\ 
is \XCOa\ $= 2\times 10^{20}~{\rm H}_2~\pcc$ (K-\kms)$^{-1}$ and
the right-most column of Table 2 shows the predicted CO brightness 
derived from N(\HH) as \WCOa\ = N(\HH)/\XCOa.  
The calculated values of \WCOa\ for J1723 
and J0942 from the entries in Table 2 are 2.8 and 2.0 K-\kms, 
respectively (the observed value for J1723 is 2.4 K-\kms), and 
there are 3 other cases in the higher-DNM group for which \WCOa\ 
$\ga 1.5$ K.  Thus, a CO survey with a robust detection limit of 1 K-\kms\ 
over an approximately 8 \kms\ velocity interval (the range observed here) 
might detect CO in a few other sightlines in the group with higher 
N(DNM). 

\subsection{The CO-\HH\ conversion factor}

A direct comparison of the measured \WCO\ and inferred N(\HH) toward 
J1723 gives 
N(\HH)/\WCO\ $ = 2.3\times 10^{20}~{\rm H}_2~\pcc$ (K-\kms)$^{-1}$:
the \hcop\ abundance chosen here quite generally yields 
\XCO\ $\approx$ \XCOa\ in direct comparison between \hcop\ absorption 
and CO emission in galactic disk gas \citep{LisPet+10,GerLis17}.  
By contrast, studies like that of 
\cite{Pla15Cham} using dust and $\gamma$-ray or dust and X-ray 
\citep{LalSno+16} measures of column density consistently find small 
values \XCO\ $ \la 0.7 \times 10^{20}~{\rm H}_2~\pcc$ (K-\kms)$^{-1}$
when averaged over whole structures or cloud complexes. If such small values 
of the CO-\HH\ conversion factor were applied to the sightlines detected
here in \hcop, the expected 
\WCO\ would have been three times higher than those in the right-most 
column of Table 2, and most of the sightlines in the higher DNM group 
would have shown detectable CO emission, some with \WCO\ $\ge 5$ K-\kms.  
These were not seen by the NANTEN observations at 
4\arcmin\ spatial resolution and also are not present in the Planck Type 2 
CO map at 6\arcmin\ resolution.

For this reason we use \XCO\ = \XCOa\ for the weakly-emitting diffuse 
molecular gas detected in the outskirts of Chamaeleon, while noting 
that smaller values apply in gas that is brighter in CO. Small regions 
of high CO brightness are in fact seen in the immediate vicinity of 
sightlines selected on the basis of their \hcop\ absorption when mapped 
in emission at 1\arcmin\ resolution with the ARO KP 12m antenna 
\citep{LisPet12}.  Moreover, the CO-\HH\ conversion factors 
N(\HH)/\WCO\ averaged over the CO hotspots observed by \cite{LisPet12} 
are smaller than \XCOa, as is expected for brighter CO emission 
from diffuse molecular gas \citep{Lis17CO}. 

\subsection{Molecular, missing, and CO-dark gas}

As shown in Table 2, somewhat more molecular gas was detected in total
than was actually required by the DNM analysis of \cite{Pla15Cham}.  Fig.
6 illustrates for the individual sightlines the manner in which the 
molecular gas absorption acts to make up the gas deficit.
Along the horizontal axis is the total gas measure that is globally
compared with and fit to the Fermi gamma ray emission and Planck dust 
optical depth maps to derive N(DNM), namely,  
N(H) = N(H I)$|_{\rm cham}$+N(DNM)+2\XCO\WCO\ with \XCO\ = \XCOa\
and \WCO\ = 0 for all sources except J1723.  Along the vertical axis
we show two different versions of the total column density N(H) that
is directly detectable in emission and/or absorption: at top is the 
combination of N(H I) and \HH\ determined from CO emission 
N(H) = N(H I)$|_{\rm cham}$+2\XCO\WCO\
while the lower panel shows  N(H) represented as the combination of 
N(H I) with \HH\ determined in  \hcop\ absorption, 
N(H) = N(H I)$|_{\rm cham}$+2W$_\hcop/3\times10^{-9}$.
The error bars assume $\pm50\%$ errors in \WCO\ and N(\hcop)/N(\HH).

In Fig. 6, the 5 sightlines with small N(DNM) are on the equality-line
in the upper panel when the CO emission is used, and they remain near 
it when \HH\ is derived from \hcop\ because they intercept mostly
atomic gas with only very weak molecular absorption.  The sightline 
toward J1723 is near the equality line when CO 
is used and somewhat closer to it when N(\HH) is derived from \hcop\
for reasons that are discussed in Sect 3.5.  The remaining 7 sightlines
lie well below the equality line when N(\HH) is derived from CO emission
in the upper panel and bracket it when \hcop\ is used.

\subsection{Saturation of the $\lambda$21cm emission}

\cite{FukTor+15} and \cite{OkaYam+17} have argued that CO emission does not miss 
any substantial amount of molecular gas and that high optical depth and 
saturation of the $\lambda$21cm H I profile can account for the DNM that 
is seen in analyses like that of \cite{Pla15Cham} and \cite{RemGre+18}.  
Given the pervasive 
molecular absorption discovered in this work, it seems difficult to 
support any argument that dismisses the importance of diffuse molecular gas 
that is missed in CO emission.  However, the possible saturation 
of H I emission on some lines of sight is a somewhat separate issue.  
The DNM analysis was performed for a series of uniform gas temperatures 
ranging from 125 K to 800 K, and it was concluded that the optically thin 
limit provided the best overall fit to interpret H I emission.  However, 
it is  possible that the amount of atomic gas is underestimated in
localised regions. 

With reference to Table 2, there is a shortfall of molecular gas, 
2N(\HH) $<$ N(DNM)/2, for three of the four sources having the highest 
DNM values.  These are the sightlines at intermediate N(H) that fall 
below the equality line in the lower panel of Fig. 6 and two of them 
have quite small  values of N(H)/\EBV\ in Table 2. Our results show 
that the loss of gas due to saturation of the H I line profile is small 
overall because the amount of gas that would need to be added to those
sightlines to bring their N(H) or N(H)/\EBV\ 
in line with the other sources is a small fraction of the total amount 
of gas sampled.  That said, saturation of the H I profile may occur 
along three of the sightlines studied here, and H I absorption measurements 
in those directions might shed some light on this issue.

\begin{table*}
\caption[]{Sightline and spectral line properties}
{
\small
\begin{tabular}{lcccccccccc}
\hline
Source&$\alpha$(J2000)&$\delta$(J2000)&$l$&$b$&\EBV$^a$&N(H I)$^b$&$S_{89.2}$&${\sigma_{l/c}}^c$&{W$_\hcop$}$^d$&{W$_{\rm HCN}$}$^e$ \\
 & hh.mmssss & dd.mmssss &\degr & \degr & mag & $10^{20}\pcc$ & Jy & & \kms  & \kms \\ 
\hline
J0942-7731&09.424275 &-77.311158 &293.321 &-18.329 &0.33&9.1&0.203 &0.0400&1.142(0.067)$^f$ & $<$ 0.155\\ 
J1058-8003&10.584331 &-80.035416 &298.010 &-18.288 &0.15&6.0&1.189 &0.0067&0.201(0.009)& $<$0.032\\
J1136-6827&11.360210 &-68.270609 &296.070 &-6.590  &0.47&21.7&0.466&0.0195&1.241(0.035)& 0.293(0.008)\\
J1145-6954&11.455362 &-69.540179 &297.316 &-7.747  &0.38&16.8&0.537&0.0188&0.870(0.031)& 0.150(0.007)\\
J1147-6753&11.473340 &-67.534176 &296.958 &-5.767  &0.30&23.0&1.552&0.0064&0.054(0.010)& $<$0.034\\
J1152-8344&11.525322 &-83.440943 &301.238 &-21.058 &0.28&9.8&0.113 &0.0719&0.240(0.070)& $<$0.342\\
J1224-8313&12.245438 &-83.131010 &302.095 &-20.391 &0.26&8.7&0.300 &0.0249&0.945(0.044)& 0.345(0.040)\\ 
B1251-7138&12.545983 &-71.381840 &303.213 &-8.769  &0.28&17.0&0.574&0.0166&0.139(0.027)& $<$0.091\\
J1312-7724&13.123874 &-77.241306 &304.122 &-14.582 &0.46&11.0&0.207 &0.0390&0.271(0.036)& $<$0.202\\
J1550-8258&15.505916 &-82.580650 &308.272 &-22.047 &0.11&6.3&0.352 &0.0261&0.242(0.040)& $<$0.158\\
J1617-7717&16.174928 &-77.171846 &313.426 &-18.854 &0.09&6.2&2.464 &0.0044&0.059(0.007)& $<$0.022\\
J1723-7713&17.235085 &-77.135020 &315.688 &-21.800 &0.26&7.7&0.309 &0.0322&1.507(0.052)& 0.952(0.030)\\
J1733-7935&17.334070 &-79.355537 &313.606 &-23.268 &0.14&7.3&0.597 &0.0149&0.057(0.020)& $<$ 0.078\\
\hline
\end{tabular}}
\\
$^a$ \cite{SchFin+98} \\
$^b$ N(H I) $= \int {\rm T}_{\rm B}~{\rm dv} \times 1.823 \times 10^{18}\pcc $ from the Gass III H I profile \citep{KalHau15} \\
$^c$ line/continuum rms at 89.2 GHz at zero optical depth \\
$^d$ N(\hcop) $= 1.10 \times 10^{12}\pcc~{\rm W}_\hcop$ \\
$^e$ N(HCN) $= 1.89 \times 10^{12}\pcc~{\rm W}_{\rm HCN}$. Upper limits are $3\sigma$ \\
$^f$ Quantities in parenthesis are the standard deviation \\
\end{table*}


\begin{table*}
\caption[]{Target by target sightline gas and dust properties in descending N(DNM) order}
{
\small
\begin{tabular}{llcccccccc}
\hline
N(DNM)&Source&\EBV&{N(H I)$|_{\rm tot}$}$^a$&{N(H I)$|_{\rm cham}$}$^b$&N(DNM)$^c$&2N(\HH)$^d$&N(H)/\EBV$^e$&N(\HH)/\XCOa$^f$ \\
 $10^{20}\pcc$  &   & mag  &  $10^{20}\pcc$  &  $10^{20}\pcc$  &  $10^{20}\pcc$   &  $10^{20}\pcc$   & $10^{21}\pcc$/mag & K-\kms  \\
\hline
$\ga2$ & J1152$^g$   & 0.28 &9.8 &5.05 & 5.15 & 1.76 & 4.1 & 0.5 \\
 & J1312             & 0.46 &11.0 & 6.71 & 3.96 & 1.98 & 2.8 & 0.5 \\
 & J0942$^g$         & 0.33 &9.1 & 6.17 & 3.82 & 8.20 & 5.2  & 2.0 \\
 & J1058             & 0.15 & 6.0 & 4.70 & 3.71 & 1.50 & 5.0 & 0.4 \\
 & J1224$^{g,h}$     & 0.26 & 8.7 & 5.63 & 3.35 & 7.20 & 6.1 & 1.8 \\
 & J1145$^{g,h}$     & 0.38 & 16.8 & 9.98 & 2.23  & 6.20 & 6.1 & 1.5 \\
 & J1723$^{g,h,i}$   & 0.26 &7.7 & 5.00& 1.95  & 11.1 & 7.2 & 2.8 \\
 & J1136$^{g,h}$     & 0.47 &21.7 &  9.15   & 1.72  &5.9 & 7.0   & 1.6 \\
\hline
&mean($\sigma$)& 0.33(0.11) &12.7(6.3) & 6.6(2.0)&3.3(1.3)&5.6(3.6) & 5.4(1.4) &1.4(0.9) \\
\hline
$\la 0.5$ & J1617 & 0.09 &6.2 & 4.31  & 0.45  & 0.43 & 7.3 & 0.1 \\
          & J1550 & 0.11 &6.3 & 4.16  & 0.27  & 1.90 & 7.4 & 0.5 \\
          & J1147     & 0.30 &23.0 & 7.73 & 0.00  & 0.40 & 7.8 & 0.1 \\
          & B1251$^g$ & 0.28 & 17.0 & 6.60  & 0.00  & 0.84 &6.0 & 0.2 \\
          & J1733$^g$ & 0.14 & 7.3 &5.25  & 0.00  & 0.58 & 5.2 & 0.2 \\
\hline
&mean($\sigma$)& 0.18(0.10) &12.7(8.6)  &5.6(1.5)& 0.14(0.21) & 0.83(0.62) & 6.7(1.1) & 0.22(0.16) \\
\hline
\end{tabular}}
\\
$^a$ As in Table 1 \\
$^b$ N(H I)$|_{\rm cham}$ is N(H I) associated with the Chamaeleon complex \\
$^c$ N(DNM) from \cite{Pla15Cham}\\
$^d$ N(\HH) = N(\hcop)$/3\times10^{-9}$\\
$^e$ N(H) = 2N(\HH)+N(H I)$|_{\rm tot}$  \\
$^f$ The predicted integrated CO J=1-0 brightness (\WCO) for X$_{\rm CO} = 2\times 10^{20}~{\rm H}_2 \pcc$ (K-\kms)$^{-1}$ \\
$^g$ \cch\ detected \\
$^h$ HCN detected \\
$^i$ \WCO\ = 2.4 K-\kms \\
\end{table*}

\begin{table*}
\caption[]{Mean column densities according to sightline sample$^{a,b}$}
{
\small
\begin{tabular}{lcccc}
\hline
Sample & $<$N(DNM)$>$ & $<$2N(\HH)$>$ & $<$N(H I)$|_{\rm cham}>$ & $<$2N(\HH)$>$/$<$N(H)$>$$^c$  \\
\hline 
all & 2.1(1.9)        & 3.8(3.7) & 6.2(1.8) & 0.38 \\
high DNM & 3.3(1.3)   & 5.6(3.6) & 6.6(2.0) & 0.46 \\  
low DNM  & 0.14(0.21) & 0.8(0.6) & 5.6(1.5))& 0.13 \\
\hline
\end{tabular}}
\\
$^a$ All column densities are in units of $10^{20}\pcc$ \\
$^b$ Quantities in parentheses are the standard deviation \\
$^c$ N(H) =  N(H I)$|_{\rm cham} + 2$N(\HH) \\
\end{table*}

\section{Summary}

Using the ALMA telescope we searched for and detected 89.2 GHz \hcop\ 
absorption in 13 directions lying toward the outskirts of the Chamaeleon 
cloud complex as summarized in Table 1 and illustrated on the sky in
Fig. 1.   The directions were selected to include the known mm-wave 
calibrator sources with fluxes above 125 mJy in March 2016 when the 
observing proposal was written, and a larger complement of strong
calibrators has not been forthcoming in the intervening time.  
CO emission and \hcop\ absorption profiles are compared in Fig. 2: 
CO emission had been firmly detected in only one of these directions, toward 
the source J1723.  Weak \cch\ absorption was detected in 8 directions and 
HCN emission in 4 directions.

As shown in Table 2, the sightlines fall into two groups having eight 
and five members, depending on whether the column density of dark neutral 
medium  N(NDM) $\ga 2.0\times 10^{20}\pcc$ or N(DNM) $< 0.5\times 10^{20}\pcc$,
respectively. The two groups have similar mean total H I column density 
$<$N(H I)$|_{\rm tot}> = 13\times 10^{20}\pcc$, comparable N(H I) associated 
with the Chamaeleon complex $<$N(H I)$|_{\rm cham}> = 5.6-6.6\times 10^{20}\pcc$ (smaller
in the low-DNM group) and mean total column density per unit reddening
$<$N(H)/\EBV $>$ $\approx 5.4-6.7 \times 10^{21} \pcc$ (mag)$^{-1}$. By contrast,
the group of eight sightlines with more DNM has much larger mean reddening 
$<$\EBV$>$ = 0.33 vs. 0.18 mag, mean DNM column density
$<$N(DNM)$>$ = 3.3 vs. 0.14 $\times 10^{20} \pcc$ and mean column density of 
H-nuclei in molecular form $<$2N(\HH)$>$ = 5.6 vs 0.83 $\times 10^{20} \pcc$. 

As shown in Figs. 3 - 5 and discussed in Sect. 3.2 there is a kinematic 
signature to the
difference between the two groups, as the group with higher N(DNM) has stronger
H I emission at 2 \kms\ $\la$ v $\la$ 8 \kms, where the  bulk of the molecular 
absorption also occurs.  Molecular absorption in the group having smaller
N(DNM) is confined to velocities v $\la 2$ \kms\ even though the sightlines
comprising the group span the entire breadth of the cloud complex (Fig. 1)
and in principle could have sampled the full range of velocities.  Molecule 
formation in the group with higher DNM may have been enhanced by slow shocks 
associated with turbulence.
 
On the whole, somewhat more \HH\ was detected in \hcop\ than previously found 
in the DNM analysis (Table 2, Sect. 3.3) but the neutral gas in the outskirts 
of Chamaeleon 
is mostly atomic.  \HH\ bears 46\% of the H-nuclei in the group having higher 
N(DNM), 13\% in the gas having lower N(DNM) and 36\% in total.  This is within 
the range of  estimates of the molecular gas fraction in the diffuse ISM as a
whole as noted in Sect. 3.3.  There  is no reservoir of ``dark'' molecular gas 
that went undetected as part of the inventory of dark neutral medium.  Saturation 
of the H I profile may have caused some atomic gas to be missed along 3 of the 
13 sightlines observed here (Sect. 3.6), and observations of HI absorption
might help to understand why the molecular gas column density was smaller 
than that of the inferred DNM in those directions.  Nonetheless, DNM is 
overwhelmingly in the form of molecular gas overall. 

Thus the situation is clear for sightlines with weak or absent CO emission in the 
outskirts of the Chamaeleon complex: the DNM is primarily molecular gas that was 
missed in CO emission and to a much lesser extent, atomic gas that might have 
been missed at $\lambda$21cm.
We note, however, that the present group of mm-wave bright quasars did not sample 
sightlines with twice-larger DNM colum densities that are present around the main 
Cha I and Cha II+III clouds. Probing the dense HI and diffuse \HH\ composition of 
the DNM in those directions would be valuable.  DNM was also detected toward regions 
with high CO intensities \WCO\ $> 7$ K-\kms, which was treated as an independent 
component in the gamma-ray analysis of \cite{RemGre+18} as it likely captures 
additional molecular gas in which  emission saturates in the main CO isotopologue
and \coth\ or \coei\ should be used.  Taking the molecular
gas census in those directions would also shed important light on \XCO\ gradients 
across molecular clouds.
\begin{acknowledgements}

This paper makes use of the following ALMA data: ADS/JAO.ALMA\#2016.1.00714.S .
ALMA is a partnership of ESO (representing its member states), NSF (USA)
and NINS (Japan), together with NRC (Canada), NSC and ASIAA (Taiwan), and
KASI (Republic of Korea), in cooperation with the Republic of Chile.  The
Joint ALMA Observatory is operated by ESO, AUI/NRAO and NAOJ.
The National Radio Astronomy Observatory is a facility of the National Science 
Foundation operated under cooperative agreement by Associated Universities, Inc. 

This work was supported by the French program “Physique et Chimie du Milieu 
Interstellaire” (PCMI) funded by the Conseil National de la Recherche Scientifique 
(CNRS) and Centre Nationa d'Etudes Spatiales (CNES).  HSL is grateful to the
hospitality of the ITU-R and the Hotel Bel Esperance in Geneva during the completion 
of this manuscript, and the incomparable Hotel Ritz in Madrid. We thank Sarah Wood,
Devaky Kunneriath and the data analysts at the North American ALMA Science Center 
for producing the imaging scripts for this project and sheparding it through 
the data reduction process.  We thank the anonymous referee for comments.

\end{acknowledgements}

\bibliographystyle{apj}

\end{document}